\documentclass[aps,twocolumn,pra,superscriptaddress,showpacs,tightenlines]{revtex4}
\usepackage{amssymb}
\usepackage{amsmath}
\usepackage{graphicx}

\begin{document}

\title{Magneto-Optical Stern-Gerlach Effect in Atomic Ensemble }
\author{Yu Guo}
\affiliation{Key Laboratory of Low-Dimensional Quantum Structures
and Quantum Control of Ministry of Education, and Department of
Physics, Hunan Normal University, Changsha 410081, China}
\affiliation{Department of Physics and Electronic Science, Changsha
University of Science and Technology, Changsha 410076, China}
\author{Lan Zhou}
\thanks{Corresponding author}
\email{zhoulan@itp.ac.cn}
\affiliation{Key Laboratory of
Low-Dimensional Quantum Structures and Quantum Control of Ministry
of Education, and Department of Physics, Hunan Normal University,
Changsha 410081, China}
\author{Le-Man Kuang}
\affiliation{Key Laboratory of Low-Dimensional Quantum Structures
and Quantum Control of Ministry of Education, and Department of
Physics, Hunan Normal University, Changsha 410081, China}
\author{C. P. Sun}
\affiliation{Institute of Theoretical Physics, Chinese Academy of
Sciences, Beijing, 100080, China}

\begin{abstract}
We study the birefringence of the quantized polarized light in a
magneto-optically manipulated atomic ensemble as a generalized
Stern-Gerlach effect of light. To explain this engineered
birefringence microscopically, we derive an effective Shr\"odinger
equation for the spatial motion of two orthogonally polarized
components, which behave as a spin with an effective magnetic
moment leading to a Stern-Gerlach split in a nonuniform magnetic
field. We show that electromagnetically-induced-transparency
mechanism can enhance the magneto-optical Stern-Gerlach effect of
light in the presence of a control field with a transverse spatial
profile and a nonuniform magnetic field.
\end{abstract}

 \maketitle

\section{Introduction}

Particles with opposite spins and non-zero magnetic moments will go
their separate ways in a nonuniform magnetic field, which is well
known as Stern-Gerlach effect~\cite{sg}. Most recently, this kind of
effect is theoretically predicted in a generalized version for some
effective nonuniform fields~\cite{cps90}, e.g., a
chirality-dependent induced gauge potential for chiral molecules
resulting from three nonuniform light fields even in the absence of
the nonuniform magnetic field~\cite{ly}.

On the other hand, a similar effect for non-polarized slow light has
been experimentally observed as the electromagnetically induced
transparency (EIT)~\cite{EIT97,EITf} enhanced deflection by a small
magnetic field gradient~\cite{Karpa}. It is also experimentally
demonstrated with a spatially distributed control
field~\cite{Scul07}. Such spatial motion of light in the EIT atomic
medium has been well explained by a semi-classical theory based on
the spatial dependence of the refraction index of the atomic
medium~\cite{ZDL} and a fully quantum approach~\cite{ZL} based on
the excitation of the dark polaritons -- the mixtures of a photon
and an atomic collective excitation~\cite{Lukin,Lukin1,scp03}. The
latter one takes advantage of revealing the wave-particle duality of
dark polaritons. And its crucial point of explanation is to derive
the effective Schr\"{o}dinger equation for the propagation of slow
light in the EIT medium.

However the above EIT-enhanced deflection of non-polarized light
can not be simply explained as an analog of the conventional
Stern-Gerlach effect since only one component of the ``spin'' is
available~\cite{Karpa,Scul07,ZDL,ZL}. An analog between light ray
and atomic beam only appears in the polarized material.
Birefringence -- the decomposition of a ray of light into two rays
dependent on the polarization when it passes through certain types
of material, is classically formalized by assigning two different
refractive indices to the material for different polarizations. In
this sense, it is the Stern-Gerlach effect of light.

In this paper we study a generalized Stern-Gerlach effect of a
quantized polarized light as a phenomenon of birefringence. The
anisotropic material is artificial. It is an atomic ensemble
controlled by a specially-designed magneto-optical manipulation
based on the EIT mechanism. Two EIT configurations are formed by an
optical field with a transverse spatial profile, since a magnetic
field removes the degeneracy of the ground-state. To represent an
analog between birefringence of quantized light and the
Stern-Gerlach effect, an effective Schr\"{o}dinger equation is
established for the spatial motion of two polarized components of
light. Such effective equation of motion describes a quasi-spin with
an effective magnetic moment in an effective nonuniform magnetic
field. The spatial gradient results from the transverse spatial
profile of the optical field, and the effective magnetic moment is
proportional to the two photon detuning with connection to the
corresponding optical field and the atomic transition.

This paper is organized as follows: In Sec.~\ref{sec:two}, we
present the theoretical model for four-level atoms with a tripod
configuration in the presence of nonuniform external fields, and
give an analytical solution of the Heisenberg equations of this
atomic ensemble in the atomic linear response with respect to the
probe field. In Sec.~\ref{sec:three}, an effective Schr\"{o}dinger
equation is derived for the spatial motion of two orthogonally
polarized components, which behave as a spin with an effective
magnetic moment. In Sec.~\ref{sec:four}, the symmetric and
asymmetric Stern-Gerlach effects are investigated in the presence of
a nonuniform magnetic field with a small transverse gradient. Then
we investigate the optical Stern-Gerlach effect in
Sec.~\ref{sec:five}, which is caused by a nonuniform light field
with a Gaussian profile in the transverse direction. In
Sec.~\ref{sec:six}, we give an explanation based on dark polaritons,
which are introduced as dressed fields to describe the spatial
motion of collective excitation. We conclude our paper in the last
section.

\section{\label{sec:two}Magneto-Optically Controlled Atomic Ensemble}

\begin{figure}[tbp]
\includegraphics[width=8 cm]{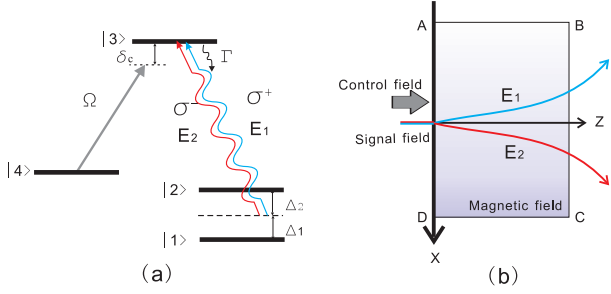}
\caption{\textit{(Color online)} (a)Energy diagram of atoms
interacting with a coupling field and a linear-polarized probe field
in the presence of a magnetic field parallel to the field
propagation direction. $\Omega $ is the Rabi frequency. (b)
Schematic diagram of light deflection in the atomic medium.}
\label{fig:1}
\end{figure}

The system we consider is in a gas-cell ABCD shown in Fig.
\ref{fig:1}(b). It is an ensemble of $2N$ identical and
noninteracting atoms with a tripod configuration~\cite{pet} of
energy levels labeled as $\left\vert i\right\rangle $ ($i=1,2,3,4)$,
see Fig. \ref{fig:1}(a).
Here, the sub-manifold is spanned by two Zeeman levels $\left\vert
1\right\rangle $ and $\left\vert 2\right\rangle $. The energy levels are
shifted by the corresponding amount $\Delta _{i}=\mu _{i}B,$ which is
determined by the applied magnetic field along $z$-axis. Here, the magnetic
moments $\mu _{i}=m_{F}^{i}g_{F}^{i}\mu _{B}$ are defined by the Bohr
magneton $\mu _{B}$, the gyromagnetic factor $g_{F}^{i}$ and the magnetic
quantum number $m_{F}^{i}$ of the corresponding state $\left\vert
i\right\rangle $. The excited state $\left\vert 3\right\rangle $ with $%
m_{F}^{3}=0$ is coupled to state $\left\vert 1\right\rangle $ ($\left\vert
2\right\rangle $) with $m_{F}^{1}=-1$ ($m_{F}^{2}=1$) via a $\sigma ^{+}$ ($%
\sigma ^{-}$) component $E_{1}$($E_{2}$) of the linear-polarized probe field
with frequency $\nu $ and wave vector $\mathbf{k}=k\hat{e}_{z}$. A classical
control field with frequency $\nu _{c}$ and wave vector $\mathbf{k}_{c}=k_{c}%
\hat{e}_{z}$ drives the atomic transition $\left\vert 3\right\rangle $-$%
\left\vert 4\right\rangle $ with spatially dependent Rabi frequency $\Omega
\left( \mathbf{r}\right) $.

The Hamiltonian of the total system $H=H^{(A)}+H^{(F)}+H^{(I)}$ is
written in terms of the collective atomic operators
$\tilde{\sigma}_{\mu \nu }\left( \mathbf{r},t\right)
=(1/N_{r})\sum_{r_{j}\in N_{r}}\tilde{\sigma}_{\mu \nu }^{j}\left(
t\right) $, averaged over a small but macroscopic volume containing
many atoms $N_r=(2N/V)dV\gg1$ around position $\mathbf{r}$, where
$2N$ is the total number of atoms and $V$ is the volume of the
medium \cite {Lukin,Lukin1}, and $\tilde{\sigma}_{\mu \nu
}^{j}\left( t\right) =\left\vert \mu \right\rangle _{j}\langle \nu
|$. Here, $H^{(F)}$ is the free Hamiltonian of the radiation field.
Neglecting the kinetic term of atoms, Hamiltonian of atomic part
reads
\begin{eqnarray}
\label{1}
H^{(A)} &=&\frac{2N}{V}\int d^{3}r\left[ \left( \omega
_{0}+\Delta _{1}\right) \tilde{\sigma}_{11}+\left( \omega
_{0}+\Delta _{2}\right) \tilde{
\sigma}_{22}\right.  \notag \\
&&\left. +\omega _{3}\tilde{\sigma}_{33}+\left( \omega _{4}+\Delta
_{4}\right) \tilde{\sigma}_{44}\right] ,  \label{SHEL-02}
\end{eqnarray}
where $\omega _{1}=\omega _{2}=\omega _{0}$, and $\omega
_{\mu},(\mu=1,2,3,4)$ are the atomic energy level spacing in the
absence of the magnetic field. Under the electric-dipole
approximation and the rotating-wave approximation, the interaction
between the atomic ensemble and the electromagnetic fields reads
\cite{ZL,Lukin,Lukin1}
\begin{eqnarray}
\label{2} H^{(I)} &=&-\frac{2N}{V}\int d^{3}r\left[\Omega
\tilde{\sigma}_{34}e^{i(k_{c}z-\nu
_{c}t)}+d_{31}\tilde{E}_{1}^{+}\tilde{\sigma}
_{31}\right.  \notag \\
&&\left. +d_{32}\tilde{E}_{2}^{+}\tilde{\sigma}_{32}+h.c.\right].
\end{eqnarray}
Here, $\tilde{E}_{j}^{+}$ are the positive frequency of the probe fields, $%
\Omega (\mathbf{r})$ is the Rabi frequency of the control field, which
usually depends on the spatial coordinate through the spatial profile of
driving field, and $d_{31}(d_{32})$ is the dipole matrix element between the
states $|3\rangle $ and $|1\rangle $ ($|2\rangle $).

As it is well known that EIT is a phenomenon specific to optically
thick media in which both the optical fields and the material states
are modified~\cite{EITf}, we introduce the slow varying variables
$E_{j}(\mathbf{r},t)$ for probe fields
\begin{equation}
\label{3} \tilde{E}_{j}^{+}(\mathbf{r},t)=\sqrt{\frac{\nu
}{2\varepsilon _{0}V}}E_{j}(\mathbf{r},t)e^{i(kz-\nu t)},(j=1,2)
\end{equation}
and $\sigma _{3j},(j=1,2,4)$ for the atomic ensemble
\begin{subequations}
\label{4}
\begin{eqnarray}
\tilde{\sigma}_{31} &=&\sigma _{31}\exp (-ikz)\text{,} \\
\tilde{\sigma}_{32} &=&\sigma _{32}\exp (-ikz)\text{,} \\
\tilde{\sigma}_{34} &=&\sigma _{34}\exp (-ik_{c}z)\text{.}
\end{eqnarray}
\end{subequations}
Then, the interaction Hamiltonian is rewritten as
\begin{eqnarray}
\label{5}
H_{I} &=&\frac{2N}{V}\int d^{3}r\left[ \left( \delta
_{1}\sigma _{11}+\delta
_{2}\sigma _{22}+\delta _{c}\sigma _{44}\right) \right.  \notag \\
&&\left. -\left( \Omega \sigma _{34}+gE_{1}\sigma _{31}+gE_{2}\sigma
_{32}+h.c.\right) \right],
\end{eqnarray}
in a frame rotating with respect to the probe and driving fields,
where $g=d_{31}\sqrt{v/\left( 2\varepsilon _{0}V\right)}$, which is
the same for both circular components $E_{1,2}$ due to the symmetry
of the system ($|d_{31}|=|d_{32}|$), is the atom-field coupling
constant and detunings are defined as
\begin{subequations}
\label{6}
\begin{eqnarray}
\delta _{1} &=&\omega _{0}-\omega _{3}+v+\Delta _{1}, \\
\delta _{2} &=&\omega _{0}-\omega _{3}+v+\Delta _{2}, \\
\delta _{c} &=&\omega _{4}-\omega _{3}+v_{c}+\Delta _{4}.
\end{eqnarray}
\end{subequations}
Before going on, we remind that tripod atoms are proved to be robust
systems for ``engineering'' arbitrary coherent superpositions of
atomic states \cite{tripod} using an extension of the well known
technique of stimulated Raman adiabatic passage.

The dynamics of this laser-driven atomic ensemble are described by
the Heisenberg equations
\begin{subequations}
\label{7}
\begin{eqnarray}
\dot{\sigma}_{12} &=&[i(\delta _{1}-\delta _{2})-\gamma ]\sigma
_{12}-igE_{1}\sigma _{32}+igE_{2}\sigma _{13}, \\
\dot{\sigma}_{13} &=&(i\delta _{1}-\Gamma )\sigma _{13}+igE_{1}(\sigma
_{11}-\sigma _{33})  \notag \\
&&+igE_{2}\sigma _{12}+i\Omega \sigma _{24}, \\
\dot{\sigma}_{14} &=&[i(\delta _{1}-\delta _{c})-\gamma ]\sigma
_{14}-igE_{1}\sigma _{34}+i\Omega \sigma _{13}, \\
\dot{\sigma}_{23} &=&(i\delta _{2}-\Gamma )\sigma _{23}+igE_{2}(\sigma
_{22}-\sigma _{33})  \notag \\
&&+igE_{1}\sigma _{21}+i\Omega \sigma _{24}, \\
\dot{\sigma}_{24} &=&[i(\delta _{2}-\delta _{c})-\gamma ]\sigma
_{24}-igE_{2}\sigma _{34}+i\Omega ^{\ast }\sigma _{23}, \\
\dot{\sigma}_{34} &=&-(i\delta _{c}+\Gamma )\sigma _{34}-igE_{1}^{\dagger
}\sigma _{14}+igE_{2}^{\dagger }\sigma _{24}  \notag \\
&&+i\Omega ^{\ast }(\sigma _{33}-\sigma _{44}),
\end{eqnarray}
\end{subequations}
where we have introduced the coherence relaxation rate of the ground
state $\gamma $, and the decay rate of the excited state $\Gamma $
phenomenologically. EIT is primarily concerned with the modification
of the linear and nonlinear optical properties of the probe field
perturbatively. We outline the solution of Eqs.(\ref{7}) in the low
density approximation, where the intensity of the quantum probe
field is much weaker than that of the coupling field, and the number
of photons contained in the signal pulse is much less than the
number of atoms in the sample. In the low density approximation, the
perturbation approach can be applied to the atomic part, which is
introduced in terms of perturbation expansion
\begin{equation}
\label{8} \sigma _{\mu \nu }=\sigma _{\mu \nu }^{(0)}+\lambda \sigma
_{\mu \nu }^{(1)}+\lambda ^{2}\sigma _{\mu \nu }^{(2)}+\cdots,
\end{equation}
where $\mu ,\nu =\{1,2,3,4\}$ and $\lambda $ is a continuously
varying parameter ranging from zero to unity. Here $\sigma _{\mu \nu
}^{(0)}$ is of the zeroth order in $gE_{j}$, $\sigma _{\mu \nu
}^{(1)}$ is of the first order in $gE_{j}$, and so on. By
substituting Eq.~(\ref{8}) into Eqs.~(\ref{7}) and keeping the terms
up to the first order in the probe field amplitude, the equations
for the first order atomic transition operators read
\begin{subequations}
\label{9}
\begin{eqnarray}
\dot{\sigma}_{13}^{(1)} &=&(i\delta _{1}-\Gamma )\sigma
_{13}^{(1)}+\frac{1}{
2}igE_{1}+i\Omega \sigma _{14}^{(1)}, \\
\dot{\sigma}_{14}^{(1)} &=&[i(\delta _{1}-\delta _{c})-\gamma ]\sigma
_{14}^{(1)}+i\Omega ^{\ast }\sigma _{13}^{(1)}, \\
\dot{\sigma}_{23}^{(1)} &=&(i\delta _{2}-\Gamma )\sigma
_{23}^{(1)}+\frac{1}{
2}igE_{2}+i\Omega \sigma _{24}^{(1)}, \\
\dot{\sigma}_{24}^{(1)} &=&[i(\delta _{2}-\delta _{c})-\gamma ]\sigma
_{24}^{(1)}+i\Omega ^{\ast }\sigma _{23}^{(1)}, \\
\dot{\sigma}_{34}^{(1)} &=&-(i\delta _{c}+\Gamma )\sigma _{34}^{(1)}+i\Omega
^{\ast }(\sigma _{33}^{(1)}-\sigma _{44}^{(1)}),
\end{eqnarray}
\end{subequations}
where we have assumed that the atoms are incoherently pumped in
states $|1\rangle $ and $|2\rangle $ with equal population at the
beginning. Under the adiabatic approximation that the evolution of
the atomic system is much faster than the temporal change of the
radiation field, the steady-state solutions are found
\begin{equation}
\label{10}
 \sigma _{j3}^{(1)}=\frac{gE_{j}\left( \delta _{j}-\delta
_{c}\right) }{2|\Omega |^{2}},~(j=1,2)
\end{equation}
where the condition $|\Omega |^{2}\gg \Gamma \gamma $ for the observation of
the important features of EIT is used, and we also set $\gamma =0$ for
showing the basic principle of physics.

\section{\label{sec:three}Effective Shr\"{o}dinger Equation Describing
Polarization as Spin Precession}

In this  section, we derive an effective Schr\"{o}dinger equation
for the spatial motion of two orthogonally polarized components of
light. In the linear optical response theory, the equations of
motion for the optical fields are given by~\cite{ZL}
\begin{equation}
\label{11}
 \left( i\partial _{t}+ic\partial _{z}+\frac{c}{2k}\nabla
_{T}^{2}\right) E_{j}=-2g^{\ast }N\sigma _{j3}^{(1)},
\end{equation}
which are achieved straightforwardly from the Heisenberg equations.
Here, $c$ is the velocity of light in vacuum and the transverse
Laplacian operator in the rectangular coordinates is defined as
\begin{equation}
\label{12}
 \nabla _{T}^{2}=\frac{\partial ^{2}}{\partial
x^{2}}+\frac{\partial ^{2}}{
\partial y^{2}}.
\end{equation}

Without loss of generality, we consider the propagating of probe
field confined in an $x$-$z$ plane. From Eqs.~(\ref{10})
and~(\ref{11}), the equations of motion for the probe fields read
\begin{equation}
\label{13}
i\partial _{t}E_{j}=\left[ -ic\partial _{z}-\frac{c}{2k}\frac{\partial ^{2}}{%
\partial x^{2}}+\frac{|g|^{2}N(\delta _{c}-\delta _{j})}{|\Omega |^{2}}%
\right] E_{j}.
\end{equation}
In order to write Eqs.~(\ref{13}) in a more compact form and
naturally show the superposition of the ``quasi-spin'' states of
photons, we introduce the ``spinor'' $\Phi $ and the third Pauli
matrix $S_{Z}$
\begin{equation}
\label{14}
 \Phi =\left(
\begin{array}{c}
E_{1} \\
E_{2}%
\end{array}%
\right) ,S_{Z}=\frac{1}{2}\left(
\begin{array}{cc}
1 & 0 \\
0 & -1%
\end{array}%
\right) .
\end{equation}%
Then Eqs. (\ref{13}) become
\begin{equation}
\label{15}
 i\partial _{t}\Phi =H_{e}\Phi ,
\end{equation}
which is a two-component Schr\"{o}dinger equation with the effective
Hamiltonian
\begin{equation}
\label{16}
 H_{e}=T+V(x)-\mu _{e}B_{e}(x,z)S_{Z}.
\end{equation}%
Here the effective kinetic term is given in terms of the momentum operators $P_{j}\equiv -i\partial _{j}$ ($%
j\in \left\{ x,z\right\} $),
\begin{equation}
\label{17}
 T=cP_{z}+\frac{P_{x}^{2}}{2m},
\end{equation}
which represents an anisotropic dispersion relation that, the
longitudinal motion is similar to an ultra-relativistic motion while
the transverse motion is of non-relativity with an effective mass
$m=k/c$. The scalar potential is determined by the detunings and $%
\tan ^{2}\theta $,
\begin{equation}
\label{18}
V(x)=\left[ \delta _{c}-\frac{1}{2}\left( \delta _{1}+\delta _{2}\right) %
\right] \tan ^{2}\theta,
\end{equation}%
where we have defined $\tan ^{2}\theta \equiv \left\vert
g\right\vert ^{2}N/\left\vert \Omega \right\vert ^{2}$. The
effective magnetic field $B_{e}(x,z)$ times magnetic moment $\mu
_{e}$ gives the spin-dependent potential
\begin{equation}
\label{19}
 \mu _{e}B_{e}(x,z)=(\delta _{1}-\delta _{2})\tan
^{2}\theta .
\end{equation}
Obviously, the above effective Hamiltonian totally determines the
dynamics of the probe field with quasi-spin-orbit coupling $\mu
_{e}B_{e}(x,z)S_{Z}$, which is also spatial dependent due to the
inhomogeneity of the applied field.

\section{\label{sec:four}Stern-Gerlach effects in nonuniform magnetic field}

In this section, we study the Stern-Gerlach effect of light  when
the magnetic field is inhomogeneous. Consider a linear magnetic
field $B\left( \mathbf{r}\right) =B_{0}+B_{1}x$ which is applied to
the atomic ensemble driven by a uniform classical field. Due to the
quasi-spin-orbit coupling, photons with orthogonal polarizations
separate their ways by the small transverse gradient $B_{1}$. To go
into this effect, we assume that both components of the
linearly-polarized probe beam are initially in a spatial Gaussian
state
\begin{equation}
\label{20}
 \Phi \left( x,0\right) =\frac{1}{\sqrt{\pi b^{2}}}\left[
\begin{array}{c}
1 \\
1%
\end{array}%
\right] \exp \left[ -\frac{x^{2}}{2b^{2}}-\frac{z^{2}}{2b^{2}}\right] ,
\end{equation}%
which is centered at $\left(x,z\right) =\left( 0,0\right) $ before
it enters the gas cell. Here $b$ is the width of the probe field
profile. After a period of time, from equation~(\ref{13}) it is
found that the initial Gaussian packet $\Phi \left( x,0\right) $
evolves into
\begin{equation}
\label{21}
 \Phi (x,t)=\left(
\begin{array}{c}
E_{1}(t) \\
E_{2}(t)%
\end{array}%
\right) =\left(
\begin{array}{c}
e^{-iH_{1}t}E_{1}(0) \\
e^{-iH_{2}t}E_{2}(0)%
\end{array}%
\right).
\end{equation}
Here, the effective Hamiltonian are defined by
\begin{equation}
\label{22}
 H_{j}=cP_{z}+\frac{1}{2m}P_{x}^{2}-\chi _{j}b_{0}-\chi
_{j}\zeta x,
\end{equation}
with the parameters
\begin{subequations}
\label{23}
\begin{align}
\zeta & =B_{1}\tan ^{2}\theta , \\
b_{0}& =B_{0}\tan ^{2}\theta , \\
\chi _{j}& =\mu _{j}-\mu _{4},
\end{align}
\end{subequations}
where we have assumed that the optical fields is in resonance with
the atomic transition in the absence of the external fields. By
making use of the Wei-Norman algebraic method~\cite{swna}, the wave
packet at time $t$ can be explicitly obtained
\begin{eqnarray}
\label{24}
 E_{j}\left( t\right) &=&\left( \frac{1/\pi }{b^{2}+i\frac{t}{m}}\right)
^{\frac{1}{2}} e^{i\varphi _{j}-\frac{\left( z-ct\right)
^{2}}{2b^{2}}} \nonumber \\  && \times \exp \left[-\frac{\left(
x-t^{2}\frac{\chi _{j}\zeta }{2m}\right) ^{2}( b^{2}m^2-itm
)}{2b^{4}m^2+2t^{2}}\right],
\end{eqnarray}
where we have introduced
\begin{eqnarray}
\label{25} \varphi _{j} =\chi _{j}t\left( \zeta
x+b_{0}-\frac{t^{2}}{3}\frac{\chi _{j}\zeta ^{2}}{2m}\right).
\end{eqnarray}

From Eq.~(\ref{24}) we can see that after passing through the atomic
medium with length $L$, the initial center of the probe field moves
to the position given by
\begin{equation}
\label{26} z=L,  \hspace{0.3cm}  x_{j}=\frac{\chi
_{j}B_{1}N|g|^2L^{2}}{2\Omega^2kc},
\end{equation}%
which indicate that, as long as the two Zeeman levels are not
degenerate, the trajectories of $\sigma ^{+}$- and $\sigma
^{-}$-polarized photons are bent in different directions. This means
that the initial linear-polarized beam is split into two parts with
opposite spins. Accompanying with the split of light, the
Stern-Gerlach effect along the transverse direction comes into
being.

\begin{figure}[h]
\includegraphics[width=8.5cm]{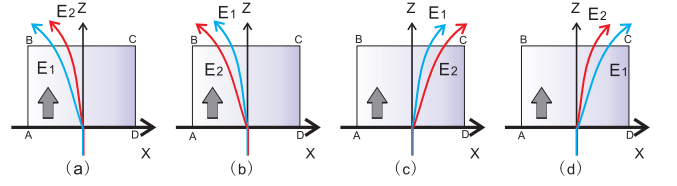}
\caption{\textit{(Color online)} Schematic illustration of the
asymmetric deflection of probe light by the nonuniform magnetic
field in conditions (a)-(d). } \label{fig:OSG2}
\end{figure}

From equation~(\ref{26}) we can see that the Stern-Gerlach effect in
the nonuniform magnetic field depends on the small magnetic field
gradient $B_1$ and the parameter $\chi _{j}$. Obviously, the
Stern-Gerlach effect disappears when the nonuniform part of the
magnetic field $B_1$ vanishes. Below we consider the split
configurations of the probe light for different $\chi _{j}$ when
$B_1>0$.  When $x_{1}=-x_{2}$, e.g., $\mu _{4}=0$ and $\mu _{1}=-\mu
_{2}$, a symmetric Stern-Gerlach effect is found -- two components
of probe beam propagate
with a mirror symmetry around $\hat{e}_{z}$-axis as shown in Fig.~\ref{fig:1}%
(b). Generally $x_{1}\neq -x_{2}$, that is $\chi _{1}\neq -\chi
_{2}$, the asymmetric Stern-Gerlach effect occurs due to detuning
mismatch, which is different from the split of atomic beam in the
magnetic field. It is the difference of $\mu _{i}$ that exerts not
identical forces on different polarized photons. The generalized
Stern-Gerlach effect of light is caused by the asymmetric gradient
force shown in Fig.2-Fig.4. It can be found that when one of the
following conditions is satisfied: (a) $\chi _{1}<\chi
_{2}<0 $; (b) $\chi _{2}<\chi _{1}<0$; (c) $\chi _{2}>\chi _{1}>0$; (d) $%
\chi _{1}>\chi _{2}>0$, two gradient forces are in the same
direction with different magnitudes for photons, therefore two
rays of light are bent in the same direction as shown in
Fig.~\ref{fig:OSG2}.

\begin{figure}[h]
\includegraphics[width=8.5cm]{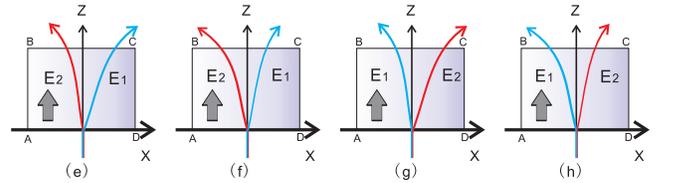}
\caption{\textit{(Color online)} Schematic illustration of the
asymmetric deflection of probe light by the nonuniform magnetic
field in conditions (e)-(h). } \label{fig:OSG3}
\end{figure}

However, when one of the following conditions is satisfied: (e) $\chi
_{1}>0>\chi _{2}$ and $|\chi _{1}|>|\chi _{2}|$; (f) $\chi _{1}>0>\chi _{2}$
and $|\chi _{2}|>|\chi _{1}|$; (g) $\chi _{2}>0>\chi _{1}$ and $|\chi
_{2}|>|\chi _{1}|$; (h) $\chi _{2}>0>\chi _{1}$ and $|\chi _{1}|>|\chi _{2}|$%
, the polarized-dependent gradient forces are in the opposite
directions with different magnitudes, therefore two rays of light
are bent in the opposite direction with different angles as shown
in Fig.~\ref{fig:OSG3}.

\begin{figure}[h]
\includegraphics[width=8.5cm]{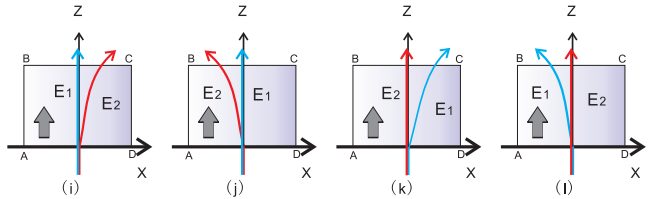}
\caption{\textit{(Color online)} Schematic illustration of the
asymmetric deflection of probe light by the nonuniform magnetic
field in conditions (i)-(l). } \label{fig:OSG4}
\end{figure}

In addition, when one of the following conditions is satisfied: (i) $\chi
_{1}=0$ while $\chi _{2}>0$; (j) $\chi _{1}=0$ while $\chi _{2}<0$; (k) $%
\chi _{2}=0$ while $\chi _{1}>0$; (l) $\chi _{2}=0$ while $\chi
_{1}<0$, only one of the two polarized photons is bent shown as
Fig.~\ref{fig:OSG4}.

\section{\label{sec:five} Stern-Gerlach effects in a nonuniform light field}

The approach developed above can also be used to investigate the
Stern-Gerlach effect of light caused by a nonuniform light field. In
this section, we turn our attention to the Stern-Gerlach effect of
slow light by the atomic ensemble driven by the optical field with a
nonuniform profile while the magnetic field is uniform.

In most experiments, the control field is continuous and has a
transverse spatial profile $\Omega \left( \textbf{r}\right) $ that
changes little in the propagating direction. To study its transverse
effects on the probe signal field, we choose the transverse spatial
profile of the control field to be Gaussian profile
\begin{equation}
\label{27}
 \Omega( \textbf{r}) =\Omega _{0}\exp \left[
-\frac{x^{2}}{2\sigma ^{2}}\right] ,
\end{equation}%
where $\sigma $ is the width of the driving field profile. We also
assume that both quasi-spins of photons are initially
linearly-polarized in a spatial Gaussian distribution
\begin{equation}
\label{28}
 \Phi \left( x,0\right) =\frac{1}{\sqrt{\pi b^{2}}}\left[
\begin{array}{c}
1 \\
1%
\end{array}%
\right] e^{-\frac{\left( x-a\right) ^{2}}{2b^{2}}-\frac{z^{2}}{2b^{2}}},
\end{equation}%
where $b$ is the width of the probe field profile and $a$ is the
initial wave-packet center of the probe field along the transverse
direction. The sign of $a$ indicates the incident position
comparatively to the left- or right-hand side of the control beam's
center $x_{0}=0$, and the magnitude $\left\vert a\right\vert $
denotes the distance from the control beam's center.

Since we concern the situation where the width of the probe field profile is
much smaller than that of the driving field profile, we expand $\left\vert
\Omega \right\vert ^{-2}$ around $a$ and retain the linear term proportional
to $x-a$,
\begin{equation}
\label{29}
 \left\vert \Omega \right\vert ^{-2}\simeq \Omega
_{0}^{-2}\left[ \exp \left(
\frac{a^{2}}{\sigma ^{2}}\right) +\frac{2a}{\sigma ^{2}}\exp \left( \frac{%
a^{2}}{\sigma ^{2}}\right) (x-a)\right] .
\end{equation}%
Then the dynamics of the probe field is governed by the unitary operators $%
U_{j}\left( t\right) =\exp \left( -iH_{cj}t\right) $, which are
generated by the effective Hamiltonians
\begin{equation}
\label{30}
 H_{cj}=cP_{z}+\frac{1}{2m}P_{x}^{2}+(\eta _{j0}+\eta
_{j1}x),
\end{equation}%
where the parameters are defined as
\begin{subequations}
\label{31}
\begin{eqnarray}
\eta _{j0}&=&-\Omega _{0}^{-2}|g|^{2}N\chi _{j}B\exp \left(
a^{2}/\sigma^{2}\right), \\
\eta _{j1}&=&-2a\Omega _{0}^{-2}|g|^{2}N\chi _{j}B\exp \left(
a^{2}/\sigma ^{2}\right)/\sigma ^{2}.
\end{eqnarray}%
\end{subequations}

\begin{figure}[tbp]
\includegraphics[width=9cm]{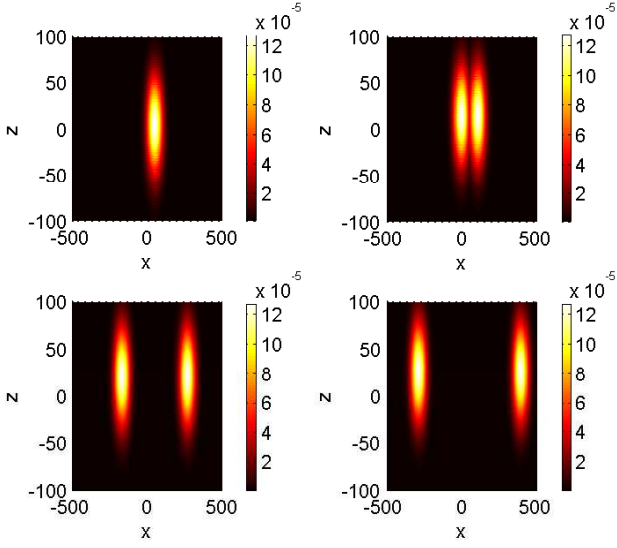}
\caption{\textit{(Color online)} The density distributions (in
arbitrary unit) of the two profiles of the probe light field for the
symmetry Stern-Gerlach effect at different times, (a) t=0, (b) t=1,
(c) t=2, (d) t=3.} \label{fig:OSG5}
\end{figure}

By using the Wei-Norman algebraic method~\cite{swna},  we solve the
time evolution problem of the probe field, and find that at time
$t$, two components of light become
\begin{eqnarray}
\label{32}
 E_{j}(t)&=&\left( \frac{1/\pi }{b^{2}+i\frac{t}{m}}\right)
^{\frac{1}{2}} e^{-i\varphi'_j -\frac{\left( z-ct\right)
^{2}}{2b^{2}}} \nonumber \\ &&\times \exp\left[-\frac{\left(
x-a-t^{2}\frac{\eta
_{j1}}{2m}\right)^{2}b^{2}m^{2}}{2b^{4}m^{2}+2t^{2}}\right],
\end{eqnarray}%
where we have introduced
\begin{eqnarray}
\label{33}
\varphi'_j &=&\left( \eta _{j0}-x\eta _{j1}\right) t+\frac{mt}{2mb^{4}+2t^{2}}-\frac{%
\eta _{j1}^{2}t^{3}}{6m}.
\end{eqnarray}
In Fig.~\ref{fig:OSG5}, we numerically demonstrate the wave packet
evolution of the two profiles of light fields.

From Eq.~(\ref{32}), it can be found that at the boundary of the
medium the center of the emergent wave function of the probe field
is changed to
\begin{eqnarray}
\label{34}
 z&=&L,  \hspace{0.3cm}
x'_{j}=a +\frac{a\chi _{j}B\exp{(a^{2}/\sigma ^{2})}|g|^{2}NL^{2}}{%
\Omega _{0}^{2}\sigma ^{2}kc},
\end{eqnarray}%
which indicates that the control field can split the
linear-polarized probe beam into two as long as the magnetic field
$B$ and $a$ are nonzero and $\chi _{1}\neq\chi _{2}$. This is a
generalized Stern-Gerlach effect in nonuniform light field. The
configurations of the splitting are completely determined by the
detuning mismatch $\chi_jB$ and the incident position $a$ of the
probe light. Comparing equation~(\ref{34}) with~(\ref{26}), we find
that an external controllable parameter $a$ is offered in this case,
which determines the splitting configurations of the probe light in
nonuniform light field. We denote $a>0$ ($a<0$) as the probe beam
shifted to the right (left) with respect to the center of control
light. Below we take $B>0$. When $a>0$, the probe beam $E_j$ bend to
the right of the axis at $x=a$ with $\chi_j>0$, while with
$\chi_j<0$, the trajectory of $E_j$ bend to the left. Actually, the
splitting around the $x=a$ axis are the same as those shown in
Figs~\ref{fig:1}(b)-\ref{fig:OSG4}(l) under the same conditions.
When $a<0$, the split around the $x=a$ axis is similar to that in
$a<0$, except exchanging the role of both components. Obviously, a
spin current transverse to the energy flow in a nonmagnetic
isotropic medium is generated by the spatial-profile of control
field.

\section{\label{sec:six}Dark-state polariton explanation}

In this section, we present a dark-state polaritons
(DSPs)~\cite{EITf,Lukin,Lukin1} explanation of the magneto-optical
Stern-Gerlach effects for slow light~\cite{slight} and show that how
the spatial motion of DSPs is associated with slow light
propagation. Originally the quasiparticle picture is introduced to
reveal the physical mechanism for the temporary transfer of
excitations to and from the medium. To put it into a mathematical
formalism, one defines the dark ($\Psi _{j}$) and bright ($\Phi
_{j}$) polariton fields \cite{EITf,Lukin,Lukin1}
\begin{subequations}
\label{35}
\begin{eqnarray}
\Psi _{j} &=&E_{j}\cos \theta -2\sqrt{N}\sigma _{j4}^{(1)}\sin \theta \text{,%
} \\
\Phi _{j} &=&E_{j}\sin \theta +2\sqrt{N}\sigma _{j4}^{(1)}\cos \theta \text{,%
}
\end{eqnarray}
\end{subequations}
which are relevant to different circular components here, they are
atomic collective excitations dressed by the quantized probe light
with its inverse relation
\begin{subequations}
\label{36}
\begin{eqnarray}
E_{j} &=&\Psi _{j}\cos \theta +\Phi _{j}\sin \theta \text{,} \\
\sigma _{j4}^{(1)} &=&\frac{1}{2\sqrt{N}}(\Phi _{j}\cos \theta -\Psi
_{j}\sin \theta )\text{.}
\end{eqnarray}
\end{subequations}

It is demonstrated in the previous references~\cite{Karpa,Scul07,ZL}
that the DSPs are matter particles with mass, momentum and magnetic
moment et.al. However, in the system under our consideration, two
dark polariton fields are excited. It is the two excited dark
polariton fields that lead to the split of the emergent light. We
will analytically show this split below. First we rewrite
Eqs.~(\ref{9}) as
\begin{subequations}
\label{37}
\begin{eqnarray}
\sigma _{13}^{(1)} &=&-\frac{i}{\Omega ^{\ast }}(\partial _{t}-d_{1})\sigma
_{14}^{(1)}, \\
\sigma _{23}^{(1)} &=&-\frac{i}{\Omega ^{\ast }}(\partial _{t}-d_{2})\sigma
_{24}^{(1)}, \\
gE_{1} &=&-2\left[ (\partial _{t}-d_{c1})\frac{1}{\Omega ^{\ast }}(\partial
_{t}-d_{1})+\Omega \right] \sigma _{14}^{(1)}, \\
gE_{2} &=&-2\left[ (\partial _{t}-d_{c2})\frac{1}{\Omega ^{\ast }}(\partial
_{t}-d_{2})+\Omega \right] \sigma _{24}^{(1)},
\end{eqnarray}
\end{subequations}
where we have defined
\begin{subequations}
\label{38}
\begin{eqnarray}
d_{1} &=&i(\delta _{1}-\delta _{c})-\gamma , \\
d_{2} &=&i(\delta _{2}-\delta _{c})-\gamma , \\
d_{c1} &=&i\delta _{1}-\Gamma , \\
d_{c2} &=&i\delta _{2}-\Gamma .
\end{eqnarray}
\end{subequations}
In terms of dark and bright polariton fields, Eqs.(\ref{11}) and
(\ref {37}) can be rewritten as
\begin{eqnarray}
\label{39}
&&\left( i\frac{\partial }{\partial t}+ic\frac{\partial }{\partial z}+\frac{c%
}{2k}\nabla _{T}^{2}\right) (\Psi _{j}\cos \theta +\Phi _{j}\sin \theta )=
\notag \\
&&i\frac{g^{\ast }\sqrt{N}}{\Omega ^{\ast }}(\partial _{t}-d_{j})(\Phi
_{j}\cos \theta -\Psi _{j}\sin \theta ),
\end{eqnarray}%
and
\begin{eqnarray}
\label{40}
&&g\sqrt{N}(\Psi _{j}\cos \theta +\Phi _{j}\sin \theta )=  \notag \\
&&-\left[ (\partial _{t}-d_{cj})\frac{1}{\Omega ^{\ast }}(\partial
_{t}-d_{1})+\Omega \right]   \notag \\
&&\times (\Phi _{j}\cos \theta -\Psi _{j}\sin \theta ).
\end{eqnarray}%

Under conditions of EIT, i.e., for negligible absorption, the excitation of
bright polariton field $\Phi _{j}$ vanishes approximately. Then the dynamics
of the dark polariton fields $\Psi _{j}$ are governed by the Schr\"{o}%
dinger-like equations~\cite{ZL}
\begin{equation}
\label{41} i\partial _{t}\Psi _{j}=[T_{B}+V_{Bj}\left(
\textbf{r}\right) ]\Psi _{j},
\end{equation}
where the polarized-dependent effective potentials is given by
\begin{equation}
\label{42} V_{Bj}\left( \textbf{r}\right)=-\chi
_{j}B(\mathbf{r})\sin ^{2}\theta,
\end{equation}%
which are induced by the atomic response to the external
spatial-dependent field. And the effective kinetic operator is
defined by
\begin{equation}
\label{43} T_{B}=v_{g}P_{z}+\frac{P_{x}^{2}}{2m_{B}},
\end{equation}%
which also shows a similar anisotropic dispersion to that mentioned
above. However, besides the effective mass $m_{B}=k/v_{g}$, the
velocity $v_{g}=c\cos ^{2}\theta $ along $z$-direction can be
controlled by the amplitude of the control field. By adiabatically
rotating the mixing angle $\theta $ from $0$ to $\pi /2$, the
polariton is decelerated to a full stop, thus all the information
carried by different degrees of freedom of the probe field is stored
in the atomic medium.

In the linear magnetic field $B\left( \mathbf{r}\right) =B_{0}+B_{1}x$,
after the dark polaritons are excited in the atomic medium with Gaussian
distribution
\begin{equation}
\label{44}
\Psi _{j}\left( 0\right) =\frac{1}{\sqrt{\pi b^{2}}}\exp \left[ -\frac{x^{2}%
}{2b^{2}}-\frac{z^{2}}{2b^{2}}\right] \text{,}
\end{equation}
we find that these dark polaritons achieve different transverse
velocities
\begin{equation}
\label{45}
 v_{jx}=\frac{\chi _{j}L}{m_{B}v_{g}},
\end{equation}%
since the hybrid light-matter quasiparticles have different
effective magnetic moments. Hence, a $\sigma ^{+}$- and a $\sigma
^{-}$-polarized beam are emergent, which are separately centered at
$\left( x_{j},z_{j}\right) =\left( \chi _{j}B_1
L^{2}\sin^2\theta/\left( 2kv_{g}\right) ,L\right) $ on the boundary
as long as a magnetic field with nonzero transverse magnetic
gradient is applied. The deflection angle of the corresponding
components of light is given by
\begin{equation}
\label{46}
 \alpha _{Bj}=\frac{v_{x}}{v_{g}}=\left[ \left( -1\right)
^{j}g_{F}^{j}-m_{F}^{4}g_{F}^{4}\right] \frac{\mu _{B}B_{1}L\tan
^{2}\theta }{kc},
\end{equation}%
which implies that a mirror symmetry splitting around $z$-axis can
be achieved if the magnetic quantum number $m_{F}^{4}=0$ can be
selected. Fig. (\ref{fig:1}b) schematically illustrates
Stern-Gerlach effect at $m_{F}^{4}=0$.

\section{\label{sec:seven}conclusions}

In summary, we have studied magneto-optical Stern-Gerlach effects
in EIT atomic ensemble. We have derived an effective Shr\"odinger
equation for the spatial motion of two orthogonally polarized
components of light. It has been shown that magneto-optical
Stern-Gerlach effects can happen in the presence of both the
nonuniform magnetic field and the nonuniform control light field.
We have also present dark-state polariton explanation of
magneto-optical Stern-Gerlach effects.

It should be pointed out that our present scheme is essentially
different from the previous work on the light deflection in atomic
medium~\cite{ZDL,ZL}. Firstly, the present scheme relies solely on
an intra-atomic process of the single-species atomic ensemble, which
causes simultaneous EIT for both components of the probe field
interacting with magnetically Zeeman split sublevels in the presence
of a driving field. The split of the optical beam is based on the
attainment of double-EIT for both components of the probe field. We
note that the double-EIT effect is not simply the sum of two
independent EIT effects, and two types of dark polaritons with
different effective magnetic moments are needed to understand
magneto-optical Stern-Gerlach effects in double-EIT atomic ensemble.
Secondly, our scheme concerns the exploitation of the two spin-state
superposition of photons, and the predicted phenomena are more
general than those in Refs.~\cite{ZDL,ZL}. Thirdly, in our scheme
the magnetic field is required for splitting the atomic sublevels,
but it is not necessary in Refs.~\cite{ZDL,ZL}.

Finally, comparing magneto-optical Stern-Gerlach effects in EIT
atomic ensemble with the spin Hall
effect~\cite{spin1,spin2,spin3,spin4} we can predict the existence
of a polarization (or quasi-spin) current, which is transverse to
the flow of photons and can be generated either by the transverse
profile of electric fields or the transverse gradient of magnetic
fields. The polarized current is carried by electrically neutral
dark-state polaritons. Because the spectral position of dark
resonances is very sensitive to magnetic fields, we do hope that
such optical analog of such ``spin Hall effect'' can be observed in
the future.

This work was supported by the NSFC with Grant No. 90203018, No. 10474104,
No. 60433050, No. 10775048, No. 10325523, and No. 10704023, and NFRPC with
Grant No. 2006CB921206, No. 2005CB724508 and No. 2007CB925204, and the
Education Department of Hunan Province.

\end{document}